\documentclass[aps,pre,twocolumn,groupedaddress]{revtex4}

\usepackage{graphicx}
\usepackage{bm}

\begin{document}

\title{
Diffusive Atomistic Dynamics of Edge Dislocations in Two Dimensions
}

\author{J. Berry}
 \email[]{berryj@physics.mcgill.ca}
\author{M. Grant}
\affiliation{
Physics Department, Rutherford Building, 3600 rue University,
McGill University, Montr\'eal, Qu\'ebec, Canada H3A 2T8
}

\author{K. R. Elder}
\affiliation{
Department of Physics, Oakland University, Rochester, MI 48309-4487
}

\date{\today}

\begin{abstract}
The fundamental dislocation processes of glide, climb, and annihilation
are studied on diffusive time scales within the framework of a continuum 
field theory, the Phase Field Crystals (PFC) model. 
Glide and climb are examined for single edge dislocations subjected to shear 
and compressive strain, respectively,
in a two dimensional hexagonal lattice. 
It is shown that the natural features of these processes are reproduced without
any explicit consideration of elasticity theory or ad hoc construction of
microscopic Peierls potentials.
Particular attention is paid to the Peierls barrier for dislocation glide/climb 
and the ensuing dynamic behavior
as functions of strain rate, temperature, and dislocation density.
It is shown that the dynamics are accurately described by simple viscous
motion equations for an overdamped point mass, where the dislocation mobility 
is the only adjustable parameter.
The critical distance for the annihilation of two edge dislocations as a 
function of separation angle is also presented.
\end{abstract}


\maketitle

\section{Introduction\label{intro}}
\paragraph*{}
Plastic flow in periodic systems is typically mediated by the motion of line
defects or dislocations. The largest challenge in developing a meaningful
theory of plasticity is often linking the microscopic behavior of individual,
discrete dislocations to the macroscopic plastic behavior of the system.
In atomic and molecular crystals for example, 
understanding the effect of dislocations on mesoscopic and macroscopic material 
properties involves the treatment of length and time scales that capture the 
relevant dynamics of individual dislocations 
($\sim$$10^{-12}s$,$\sim$$10^{-9}m$) 
through those that describe the macroscopic response of the material 
($\sim$$10^{1}s$,$\sim$$10^{-2}m$).
An important approach to the problem of spanning this large range of scales has 
been to measure the dynamics of individual dislocations and/or small numbers of 
interacting dislocations on the shortest time scales 
from Molecular Dynamics (MD) simulations 
\cite{chang99,chang02,bailey01,rodary04}.
These results are then used as input into coarse-grained, 
mesoscopic simulations 
such as Dislocation Dynamics (DD) \cite{dd90,bulatov98},
which can provide information on systems with large numbers of dislocations
under the action of experimentally accessible strains and strain rates.
\paragraph*{}
In this study, dislocation dynamics are examined on length scales comparable 
to those encountered in MD simulations, but over diffusive time scales and 
using experimentally accessible strain rates.
This approach provides a single framework that removes the vibrational 
time scales, while all of the relevant length scales can potentially be 
reached with greater 
computing power or with more advanced numerical techniques \cite{nigel05}.
In addition to atomic crystals, the results presented here may be
interpreted in terms of other periodic systems such as
Abrikosov vortex lattices in superconductors \cite{pardo98}, 
magnetic thin films \cite{sagui95,orlikowski99}, 
block copolymers \cite{harrison00}, 
oil-water systems containing surfactants \cite{laradji91}, 
and colloidal crystals.
\paragraph*{}
The PFC model describes periodic systems of a continuum field nature and 
naturally 
incorporates elastic and plastic behavior. The details of the model have been
presented elsewhere \cite{ken01}, and only the necessary equations will be 
given here. The dimensionless free energy functional is written as
\begin{equation}
F=\int{d\vec{x}\bigg[{\rho \over 2}\omega(\nabla^2)\rho+{\rho^4 \over 4}\bigg]}
\label{free}
\end{equation}
where $\rho$ is an order parameter corresponding here to density and
\begin{equation}
\omega(\nabla^2)=r+(1+\nabla^2)^2.
\end{equation}
$r$ is a
phenomenological constant related to temperature. The dynamics of $\rho$
are described by
\begin{equation}
{\partial{\rho} \over \partial{t}}=\nabla^2 \frac{\delta F}{\delta\rho}+\zeta=
\nabla^2(\omega(\nabla^2)\rho+\rho^3)+\zeta
\label{dyn}
\end{equation}
where $\zeta$ is a gaussian
random noise variable ($\langle\zeta(\vec{r}_1,t_1)\zeta(\vec{r}_2,t_2)\rangle
=D\nabla^2\delta(\vec{r}_1-\vec{r}_2)\delta(\tau_1-\tau_2)$
and $D=uk_BTq_{0}^{d-4}/\lambda^2$) which has
been largely neglected in this study, as will be discussed in Section III.
\paragraph*{}
In Section II, the details of how the PFC model is adapted to numerical 
simulation are outlined, and in 
Section III the simulation results for glide, climb, and annihilation 
are presented and analyzed.
Section IV includes a summary, comparison with other recent phase field 
simulations of dislocations, and discussion of further developments.

\section{Simulation Method\label{section2}}

\subsection{Discretization, Initial Conditions, and Boundary Conditions
\label{section2a}}
\begin{figure}
\includegraphics[width=65mm]{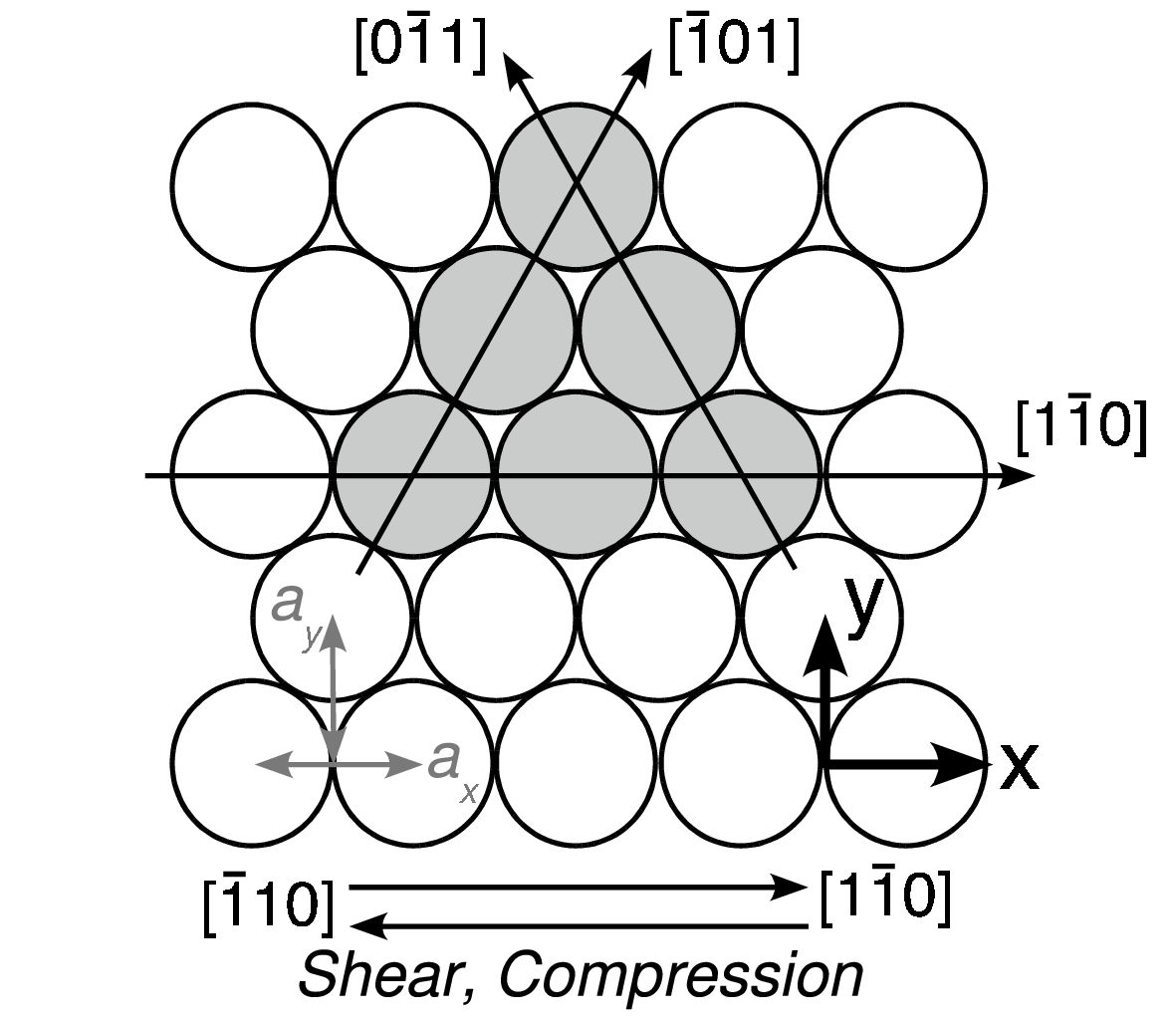}
\caption{
Schematic of (111) plane in a FCC crystal corresponding to the 2D system 
of interest.
\label{hex}
}
\end{figure}

\paragraph*{}
Eq.\ (\ref{dyn}) was solved numerically in two dimensions using the 
'spherical laplacian' approximation for $\nabla^2$ \cite{oono_puri}
and a forward Euler discretization for the time derivative.
Periodic boundary conditions were applied in all directions for glide 
simulations and mirror boundary conditions were used perpendicular to the
climb direction in climb simulations. 
To create a system with a single edge
dislocation, an initial condition consisting of a hexagonal one-mode solution 
for $\rho(x,y)$
was applied with N atoms/row in the lower half and N+1 atoms/row in the upper
half. The hexagonal state is expressed analytically as 
\begin{equation}
\rho(x,y)=A\bigg[\cos{(qx)}\cos{({qy\over\sqrt{3}})}-
{1\over2}\cos{({2qy\over\sqrt{3}})}\bigg]+\rho_0
\label{onemode}
\end{equation}
where
\begin{equation}
A={4\over5}\bigg(\rho_0+{1\over3}\sqrt{-15r-36\rho_0^2}\bigg),
\label{amp}
\end{equation}
$q$ is the numerically determined equilibrium wavenumber for a hexagonal
state at a given value of $r$, and $\rho_0$ is the average density.
In glide simulations, the hexagonal state was
bounded at its upper and lower edges by a constant, or liquid, state 
of width approximately $4a_y$, where $a_y$ is the equilibrium lattice parameter
in the $y$-direction (Fig.~\ref{hex}). 
The same approach was used in the climb simulations, except that the liquid
was placed along the sides. 
Before applying strain, all systems were allowed to equilibrate until their 
free energy no longer changed with time.
\paragraph*{}
At a given value of $r$, the value of $\rho_0$ for the hexagonal
portion of the simulation was set to fall on the phase boundary between the 
hexagonal 
and hexagonal/constant coexistence regions. The value of $\rho_0$ for the liquid
portion of the simulation was set to fall on the boundary between the 
hexagonal/constant
coexistence region and the constant phase region. This was necessary to make the
interfaces between the hexagonal and constant phases stable, with no 
preference toward
crystallization or melting. A drawback is that this makes any comparison of 
results
at different $r$ values indirect, since $\rho_0$ must vary with $r$. 
For this reason the boundary conditions were changed when the $r$ dependence of
the dynamics was of interest. Details are discussed in the following section.

\subsection{Strain Application\label{section2b}}
\paragraph*{}
Two methods were used to apply strain to the system. In both, $\rho(x,y)$
was coupled to an external field along the outer two rows of 
particles bounding the liquid phase on each side of the system. 
This external field was set to the one-mode solution
given in Eq.\ (\ref{onemode}), and for glide (climb) was moved in 
the positive $x$-direction along the 
lower (left) rows and in the negative $x$-direction along the upper (right) 
rows, both at the same
constant velocity. The particles in the system are energetically motivated
to follow the motion of these fields, giving the effect of a physically applied
strain.
\paragraph*{}
In the first method, which will be called rigid displacement, 
Eq. (\ref{dyn}) was solved in the presence of the external fields, but
in addition,
the particles between the external fields were rigidly displaced 
along with the motion of the fields 
to ensure a linear strain profile across the width of the system. 
In the second method, this rigid displacement
was not enforced, allowing the strain profile to take whichever form 
the dynamics
of Eq.\ (\ref{dyn}) dictate. This method will be called relaxational 
displacement. In Section III, it will be shown that the dynamic 
behavior of the dislocations
can be significantly influenced by which method is used and that the
two methods may be viewed as limiting cases of rigid and diffusive response.
From this viewpoint, rigid displacement describes atomic crystals and 
relaxational displacement applies to 'softer' systems such as
colloidal crystals, superconducting vortex lattices, magnetic films, 
oil-water systems containing surfactants, and block copolymers.

\subsection{Symmetries and Time Scales\label{section2c}}
\paragraph*{}
The crystalline symmetry here is equivalent to the \{111\} family of planes in 
a FCC lattice
or the \{0001\} family of planes in a HCP lattice, for example. 
These close packed
planes and the subsequent glide directions compose the primary slip systems
in many common types of ductile, metallic crystals.
Using the FCC lattice as a reference,
application of shear in this geometry results in glide along a
$\langle110\rangle$ 
direction within a \{111\} slip plane, as shown in Fig.~\ref{hex}. 
The directions
in a HCP lattice would fall in the $\langle11\bar{2}0\rangle$ family. Climb
in this geometry was made to occur along a $\langle\bar{1}\bar{1}2\rangle$ 
direction. Shear and compression were also applied over various other rotations
as will be discussed briefly in the following.
\paragraph*{}
System sizes ranging from 676 to 56,952 particles were examined, and strain
rates ranging from 2$\times10^{-7}/t$ to 1$\times10^{-2}/t$ were used,
where $t$ is the dimensionless time introduced in Eq. (\ref{dyn}). 
These strain rates can be expressed in physical units by
matching the time scales of the model to those of typical metals near their 
respective melting temperatures. This is done by equating vacancy diffusion 
constants, $D_v$, which have been calculated analytically for this model in 
\cite{ken04}, and which may range from $10^{-8}$--$10^{-13}cm^2/s$ for typical 
metals \cite{lubensky85}. Lattice constants, $a$, must also be equated to 
return to physical units. Using Cu at $1063^{\circ}\mathrm{C}$ as a reference 
($D_v\simeq10^{-9}cm^2/s$, $a\simeq3.61$\AA), and matching to the model at 
$r=-0.8$ ($D_v=1.78a^2/t$), the range of strain rates used converts to 
.09/$s$--4500/$s$.
Using these same parameters, the dislocation velocities observed are 
on the order of $10^{-7}$--$10^{-4}m/s$, a range well below the 
acoustic limit and accessible by experiment.
The dislocation densities 
range from approximately $10^{10}$--$10^{12}/cm^2$. 

\subsection{Simulation Output: A Preliminary Example\label{section2d}}
\begin{figure}
\includegraphics[width=88mm]{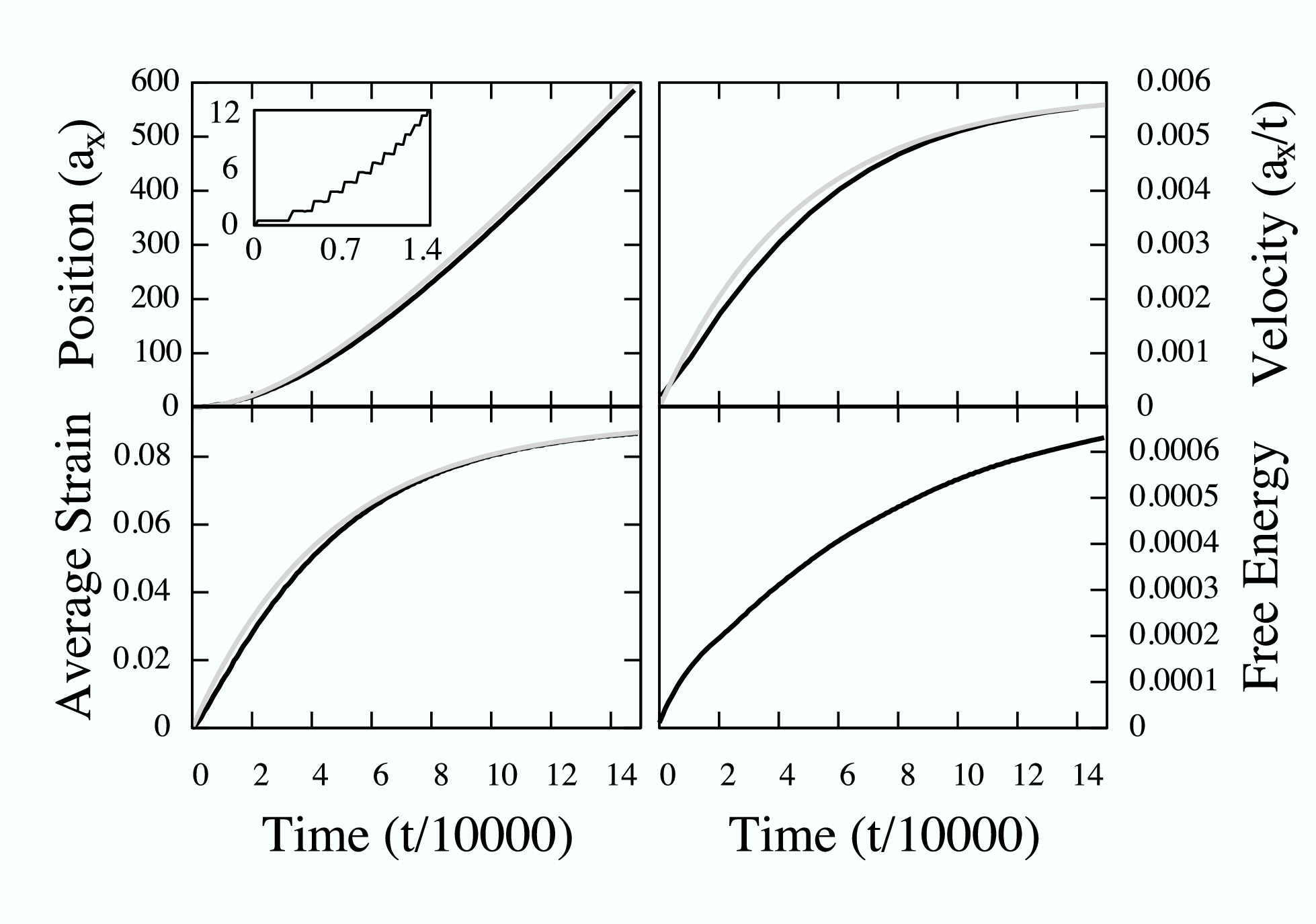}
\caption{
One set of simulation data (black lines) and the corresponding theoretical 
results (gray lines) for glide.
Parameters are $r=-0.4$, ($L_x,L_y$)=(60,56), 
and $\dot{\gamma}=2\times 10^{-6}/t$ 
The inset in the upper left corner
of the upper left plot shows a magnification of the position vs. time data at 
early times to emphasize the stick-slip nature of the motion at low velocities.
\label{gallvst}
}
\end{figure}

\begin{figure}
\includegraphics[width=88mm]{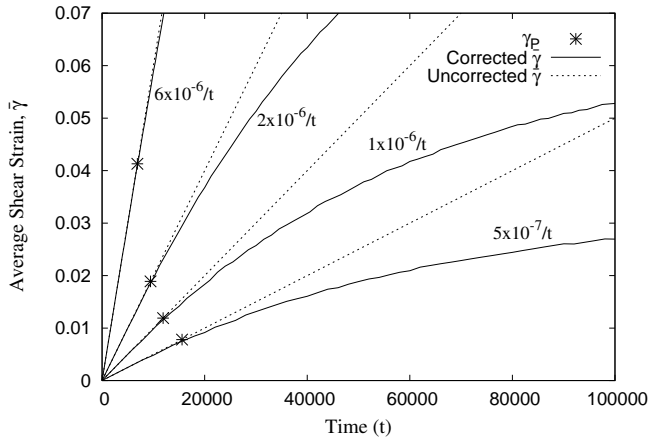}
\caption{
Corrected and uncorrected $\bar{\gamma}$ versus time at $r=-0.4$ and 
$(L_x,L_y)=(60,56)$ for various values of $\dot{\gamma}$.
The strain at which $\gamma_{P}$
is defined has been highlighted on each curve.
\label{gstrvtimepei}
}
\end{figure}

\paragraph*{}
Before presenting the analysis of all simulation data, the output from 
a single glide simulation will be presented to clarify various definitions
and results that will be of importance in interpreting the data. The
collective results from all simulations will be analyzed further in the 
following section.
\paragraph*{}
Four primary types of output were generated in each simulation, from which all
properties of interest were extracted. The variables are the instantaneous 
position and 
velocity of the dislocation, and the strain and change in free energy of the
system, all recorded as functions of time as shown in Fig. \ref{gallvst}.
The gray lines represent theoretical results which will be presented in 
Section III.
\paragraph*{}
The position was determined by locating all maxima of $\rho(x,y)$ (which can
be considered the discrete particle locations) and counting
the number of nearest neighbors for each. Any maxima with more or less than
six nearest neighbors must be near the dislocation core, and by averaging the 
positions of all maxima identified in this way, an overall dislocation 
position was inferred. The velocity was then calculated from the slope of the
position versus time data.
\paragraph*{}
The average shear strain in the system, $\bar{\gamma}$,
was measured by again locating the peaks in $\rho(x,y)$, and noting that
in equilibrium, each particle will have another
particle located a distance of $2a_y$ away in the positive $y$-direction. 
If this
particle is found to be offset some distance, $dx_i$, in the $x$-direction, 
then the 
local shear strain is equal to $dx_i/2a_y$. The average shear strain
in the system is then given by
\begin{equation}
\bar{\gamma}=\frac{1}{2Na_y}\sum_{i=1}^N dx_i
\label{strainmeas}
\end{equation}
where $N$ is the number of particles in the system.
The fourth variable, the average free energy $F$, was simply calculated from 
Eq. (\ref{free}) at regular intervals of time.
\paragraph*{}
The Peierls barrier is a measure of the resistance to 
the onset of motion in a periodic system.
In these simulations, the barrier is defined as the amount of strain that 
has been applied at the instant that
the dislocation has precessed a distance of one lattice constant. 
$\gamma_{P}$ and $\epsilon_{P}$ will denote the Peierls 
barriers for glide and climb, respectively.
For clarity, Fig.~\ref{gstrvtimepei} shows $\bar{\gamma}$ as a function of time
for a few different values of $\dot{\gamma}$. The strain corresponding to
this definition of $\gamma_{P}$ is indicated on each curve and can be seen
to correspond to the point where the measured strain begins to deviate from the
applied strain. The deviation is due to the strain relieved by the motion
of the dislocation, as will be discussed in the next section.

\section{Results and Analysis\label{section3}}

\subsection{Equilibrium Dislocation Geometry\label{section3a}}
\begin{figure}
\includegraphics[width=80mm]{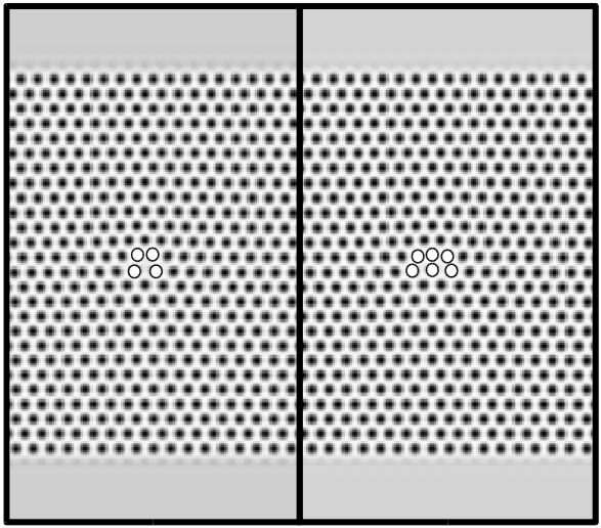}
\caption{
Stable dislocation configurations: The greyscale represents $\rho(x,y)$
and the particles around the dislocation core have been highlighted for
clarity. Left: Config. 1. Right: Config. 2.
\label{configs}
}
\end{figure}

\begin{figure}
\includegraphics[width=88mm]{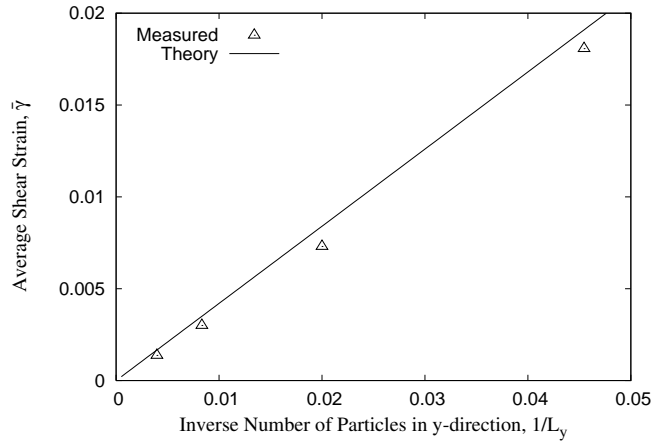}
\caption{
Equilibrium elastic strain, $\bar{\gamma}$, due to an edge dislocation plotted 
as a function of inverse system size in the $y$-direction. 
$L_x$ was fixed at 56 particles for the data shown.
\label{elasvsize}
}
\end{figure}

Following equilibration as described in the previous section, the dislocations
were found to reach one of the two stable configurations shown in 
Fig.~\ref{configs}. 
Which of the two 
configurations is selected depends sensitively on the details of the boundary
conditions as well as on the system size. Systems larger in the $x$-direction
tend to favor Config. 1, and systems larger in the $y$-direction tend to
favor Config. 2, apparently due to the greater strain relief available
at larger extensions. Systems with approximately equivalent $x$ and $y$ 
dimensions that were equilibrated with thermal noise oscillated between 
Configs. 1 and 2, indicating that the two states are approximately 
equivalent energetically. It will be shown in the following section that 
the initial configuration affects $\gamma_{P}$ but not the 
velocity of the dislocation.
\paragraph*{}
The average shear strain, $\bar{\gamma}$, in each system was measured 
and the values recorded following
equilibration have been plotted in Fig.~\ref{elasvsize} as a function of 
$1/L_y$, where $L_y$ is
the number of particles in the $y$-direction.
A simple analysis reveals that the total $\bar{\gamma}$ due to an edge 
dislocation in this geometry is equal to $\sqrt{3}b/(4L_y)$, 
where $b$ is the burger's vector of the
dislocation. This result agrees well with the measured values shown in 
Fig. \ref{elasvsize}, indicating
that the measurement technique is reliable.

\subsection{Glide: Constant Applied Shear Rate Dynamics\label{section3b}}
\paragraph*{}
Simulations were conducted using steady shear over a range of applied shear
rates ($\dot{\gamma}$),
temperatures ($r$), and system sizes ($L_x,L_y$). The dependence of the 
Peierls barrier and
the velocity vs. $\bar{\gamma}$ behavior on these
variables will be discussed in the following subsections.

\subsubsection{Peierls Barrier for Glide\label{section3b1}}
\begin{figure}
\includegraphics[width=88mm]{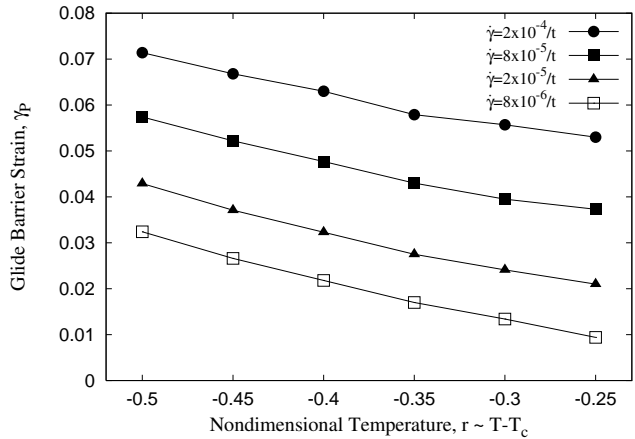}
\caption{
Temperature dependence of the Peierls strain barrier for glide 
without thermal fluctuations.
Data shown is at $\rho_0=0.25$ and ($L_x,L_y$)=(56,56) under 
rigid displacement.
\label{peivr}
}
\end{figure}

\begin{figure}
\includegraphics[width=88mm]{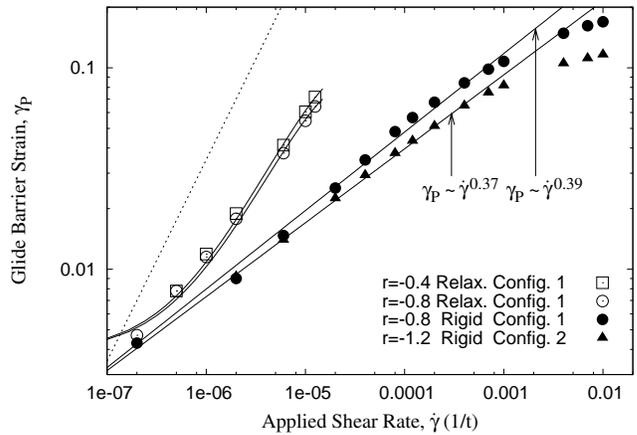}
\caption{
Measured Peierls strain barrier for glide, $\gamma_{P}$, 
as a function of applied strain rate for
the cases of rigid and relaxational displacement.
The fits to the relaxational data are from Eq. (\ref{hoptime3}) and
the fits to the rigid data are power laws as indicated in the image.
The dotted line shows a linear $\dot{\gamma}$ dependence for reference.
Note that $\gamma_{P}$ is consistently lower for Config. 2 than for
Config. 1, even with the differing $r$'s working toward the opposite effect.
\label{peivrate}
}
\end{figure}

\paragraph*{}
To test for finite size effects, $\gamma_{P}$ was measured as
a function of system size, or inverse dislocation density. Within estimated
errors, no change was observed under rigid displacement as the system size was 
increased from 676 to 56,952 particles.
Under relaxational displacement, a slight increase with
$L_y$ was noted, and is linked to the time required for the strain applied
at the edges to diffuse inward to the dislocation core. 
Diffusion is fast compared to the inverse shear rates required to apply
relaxational displacement (rows of particles slip relative to each other
at all but the lowest values of $\dot{\gamma}$), so the increase 
of $\gamma_{P}$ with $L_y$ cannot be very large. 
The nonlinear shear profile that is produced may exaggerate
this lag between the applied strain and the strain near the dislocation,
but the overall effect was nonetheless found to be relatively small.
\paragraph*{}
Next, the barrier was examined as a function of $r$, which is proportional to
the distance in temperature from $T_c$. To do this consistently, the boundary
conditions were changed to mirror rather than periodic at the top and bottom,
and the constant phase
was entirely removed from the simulation. This made it possible to vary
$r$ at a single value of $\rho_0$, isolating the temperature dependence in a 
more realistic manner. Results are shown in Fig. \ref{peivr}.
\paragraph*{}
The decrease in $\gamma_{P}$ as the melting point is approached 
is expected
since the hexagonal phase becomes less pronounced near $T_c$.
That is, $A$ decreases with increasing $r$ according to Eq. (\ref{amp}), 
and even
without thermal fluctuations a distinct temperature dependence is produced.
This decrease in $A$ corresponds to an increase in the width of
the dislocation which, according to the
Peierls-Nabarro model \cite{phillips}, lowers the Peierls barrier for glide.
With thermal fluctuations, these results did not change significantly,
though at low $\dot{\gamma}$, which is where the change would be greatest, 
it was not possible to include fluctuations and maintain reasonable 
computation times. 
Similar linear decreases in $\gamma_{P}$ as some effective 
$T_c$ is approached have been found in experiment 
\cite{wasserbach86,lachenmann70} and theory \cite{cuitino01,tang99},
along with increases in
$\gamma_{P}$ with $\dot{\gamma}$ much like those
shown in Fig. \ref{peivr}. 
\paragraph*{}
At temperatures closer to the melting point ($r\simeq -0.18$),
the dislocations began to climb at very low strains before any
glide had occurred. This is the first evidence that 
climb is the dominant process at high temperatures, as in real crystals. 
Further evidence will be presented with the climb results.
\paragraph*{}
The dependence of $\gamma_{P}$ 
on $\dot{\gamma}$ was also explicitly measured (Fig.~\ref{peivrate}). 
Both methods of shear application
result in what appears to be a power law increase in 
$\gamma_{P}$ as the shear rate is increased, where the relaxational 
displacement
data is nearly linear and the rigid displacement data appears to approach a 
limit $\gamma_{P}$ at high $\dot{\gamma}$.
\paragraph*{}
This increase is explained as follows. Extrapolating the data to
$\dot{\gamma}=0$ indicates that in all cases there is some small
strain $\gamma_{P}^0<0.5\%$ at which the dislocation will glide, given
sufficient time. Call this time from when $\gamma_{P}^0$ is reached to 
when the first glide event actually occurs $\Delta t_{hop}$.
In any given simulation, once $\bar{\gamma}>\gamma_{P}^0$, 
excess strain is being applied during the interval $\Delta t_{hop}$ 
which makes the observed $\gamma_{P}$ appear to be larger than 
$\gamma_{P}^0$. In this approximation
\begin{equation}
\gamma_{P}=\gamma_{P}^0(r)+\dot{\gamma}\Delta t_{hop}(r)
\label{hoptime1}
\end{equation}
which predicts a linear increase in $\gamma_{P}$ above $\gamma_{P}^0$. 
It is reasonable to
expect though, that $\Delta t_{hop}$ will decrease at higher strain 
rates due to the additional strain applied during the interval. In
a first approximation then
\begin{equation}
\Delta t_{hop}(r,\dot{\gamma})=\Delta t_{hop}^0(r)-
\alpha(r)\eta(d,\dot{\gamma})\dot{\gamma}
\label{hoptime2}
\end{equation}
where $\Delta t_{hop}^0(r)$ is the time required to execute a glide event
at $\dot{\gamma}=0$ and $\alpha(r)$ is a coefficient related to the
magnitude of the additional driving force applied during $\Delta t_{hop}$.
$\eta(d,\dot{\gamma})$ is a coefficient related to the efficiency of
strain transfer from the sample edges where strain is applied to the 
dislocation core, as a function of the type of displacement $d$ and 
$\dot{\gamma}$. $\eta(d,\dot{\gamma})=1$ for rigid displacement and
drops below $1$ under relaxational displacement.
\paragraph*{}
Substituting Eq. (\ref{hoptime2}) into Eq. (\ref{hoptime1}) gives
\begin{equation}
\gamma_{P}=\gamma_{P}^0(r)+\dot{\gamma}\Delta t_{hop}^0(r)-
\alpha(r)\eta(d,\dot{\gamma})\dot{\gamma}^2
\label{hoptime3}
\end{equation}
which now indicates a dependence on $\dot{\gamma}$ that must be fall
below the initial expected linear trend. And the deviation from linear
will be greatest under rigid displacement, since $\eta$ is maximum in 
this case. The data is in agreement with this expectation.
Using $\gamma_{P}^0(-0.4)=
\gamma_{P}^0(-0.8)=0.0038$, $\Delta t_{hop}(-0.4)=7277t$, 
$\Delta t_{hop}(-0.8)=6643t$, and $\alpha(-0.4)\eta=\alpha(-0.8)\eta=
1.48\times10^{-8}t^2$ produces reasonable fits to the relaxational
displacement data shown in Fig. \ref{peivrate}. 
Note that if $\Delta t_{hop}$ is negligible,
then clearly this effect will not be noticeable, but since the dynamics
are necessicarily diffusive in these simulations, it is reasonable to
expect some contribution from this effect.
\paragraph*{}
Accurate fits to the rigid 
displacement data were more difficult to obtain, most likely due to the
transition to no $\dot{\gamma}$ dependence for large $\dot{\gamma}$.
This can be shown by studying the 
evolution of $F$ under an applied shear.
In \cite{ken04}, the change in
$F$ for a one-mode approximate hexagonal solution under the action of shear
was be found by minimizing $F$ when $\rho(x,y)$ is replaced with
$\rho(x+\bar{\gamma}y,y)$. The resulting equation, valid for small
$\bar{\gamma}$, is
\begin{equation}
\Delta F_{Shear}=\frac{q^4_{eq}A^2}{6}\bar{\gamma}^2.
\label{dFrigid}
\end{equation}
In principle, this represents a rigid displacement of $\rho(x,y)$
at infinitely large $\dot{\gamma}$.
In this limit, $\gamma_{P}$ has no explicit dependence on 
$\dot{\gamma}$;
\begin{equation}
\gamma_{P}=\sqrt{\frac{6\Delta F_{P}^{Glide}(\rho_0,r)}{q^4_{eq}A^2}}.
\label{peirigid}
\end{equation}
\paragraph*{}
Dislocation dynamics in soft structures such as colloidal crystals are
reasonably expected to correspond to the case of relaxational displacement.
These systems typically exhibit very little rigidity associated with
sound modes or phonons, thus their relative softness.
Conversely, dynamics in atomic crystals are believed to correspond to the 
case of rigid 
displacement at large $\dot{\gamma}$. Atomic crystals exhibit a significant
rigidity in response to an applied shear, which can be reasonably 
approximated by a linear shear profile as is done for rigid displacement.
A more constructive way to model atomic crystals would be to explicitly
consider a phonon or wave term in the dynamics, as is being done by other
authors \cite{mpfc}.
If such modes were considered, the collective motion of 
particles in response to an applied force would naturally be enhanced, 
more resembling 
the case of rigid displacement. It is argued in this sense that the methods
of rigid displacement in the large $\dot{\gamma}$ region
and relaxational displacement represent limiting cases
of response, and that a more rigorous description including effective phonon
dynamics would fall between these limits.

\subsubsection{Atomistic Glide Mechanism\label{section3b2}}
\begin{figure}
\includegraphics[width=84mm]{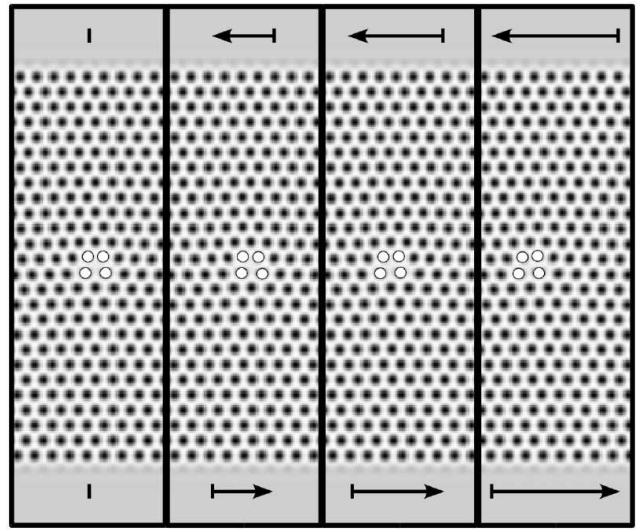}
\caption{
Atomistic glide mechanism under constant applied shear rate (the particles
around the dislocation core have been highlighted for clarity).
From left to right, $\rho(x,y)$ is shown at $t$=0, 500, 1000, and 1500,
corresponding to $\bar{\gamma}$=1$\%$, 2$\%$, 3$\%$, and 4$\%$.
The arrows indicate relative strain magnitudes and directions.
\label{glide}
}
\end{figure}

\begin{figure}
\includegraphics[width=88mm]{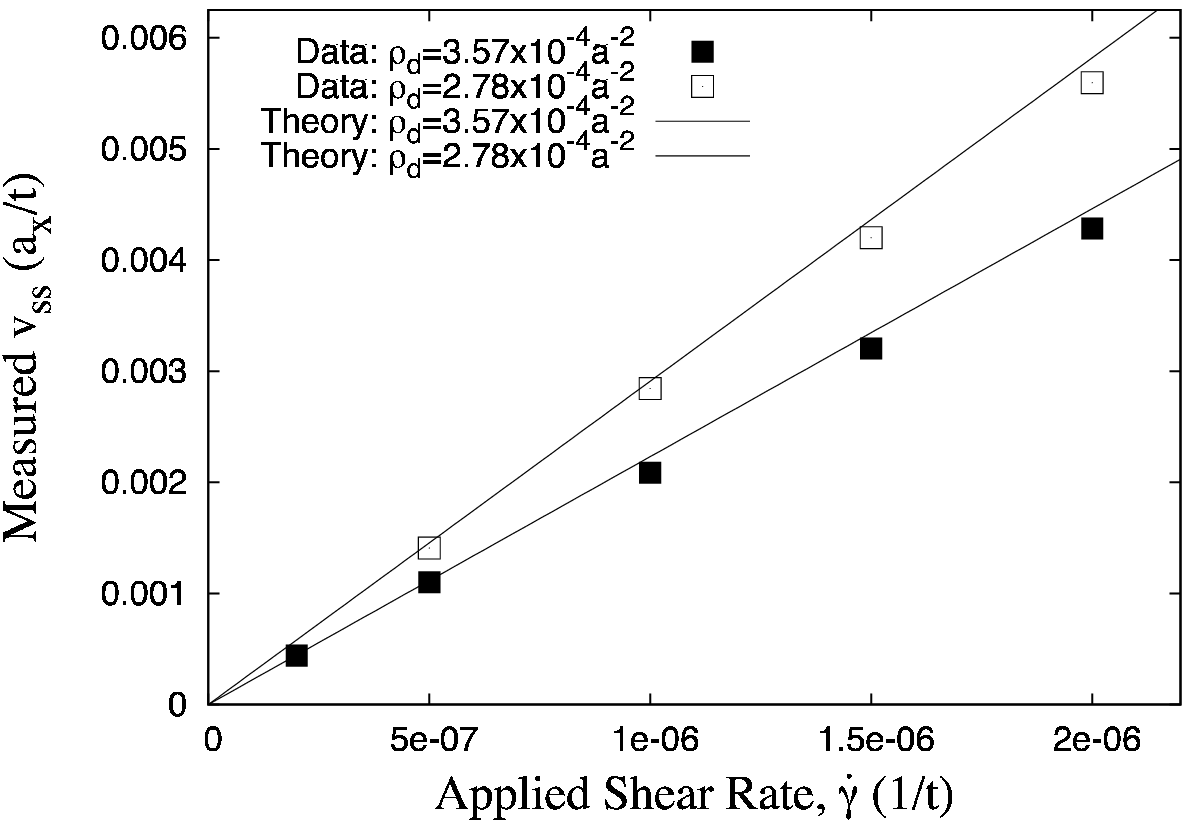}
\caption{
Measured steady-state glide velocities for two system sizes.
The upper data points are at $r=-0.4$ and the lower data points are at 
$r=-0.8$.
\label{velss}
}
\end{figure}

\paragraph*{}
The nature of the dislocation motion in these simulations 
(Fig.~\ref{glide}) is stick-slip at low velocities with
a transition to a more continuous character at high velocities. This is
expected, as the lattice barrier leads to thermally activated motion when 
$\Delta F_{Shear}$ approximately equals $\Delta F_{P}^{Glide}$, while at 
large values of
$\Delta F_{Shear}$ the barrier becomes secondary and the motion assumes
a damped character. The shear rate dictates the maximum velocity and 
therefore the extent to which the motion becomes continuous. Three regimes of
motion were observed, with selection depending on the ratio
\begin{equation}
v_{ss}=\frac{\dot{\gamma}}{\rho_d b}
\label{vss}
\end{equation}
where $\rho_d$ is the dislocation density dictated by the system size.
The reason for labeling this quantity $v_{ss}$ will soon become apparent.
\paragraph*{}
For large $v_{ss}$ ($\gtrsim0.016a/t$), the dislocation quickly 
reaches the overdamped regime and adjacent layers
of particles begin to slip relative to each other along the $x$-direction
before a steady-state velocity is achieved. Slipping usually occurs when the
strain exceeds approximately 20\% in rigid displacement or 10--15\% in 
relaxational displacement.
At moderate values of $v_{ss}$ 
($0.06\gamma_{P}\lesssim v_{ss}\lesssim0.016a/t$), the dislocation
approaches a continuous glide motion and eventually reaches a steady-state 
velocity. This velocity can be calculated by equating the Orowan equation 
\begin{equation}
\dot{\gamma}_{Plastic}=\rho_d b v
\label{orowan}
\end{equation}
to the applied shear rate, giving the quantity $v_{ss}$ defined in 
Eq.\ (\ref{vss}).
This is the glide velocity required to plastically relieve strain at exactly
the same rate at which it is being applied.
Fig.~\ref{velss} shows $v_{ss}$ versus $\dot{\gamma}$ as measured 
from simulations. The
measured values follow a linear trend as Eq. (\ref{vss})  
predicts, with the slopes in good agreement with the theoretical values.
This again shows that the plastic strain relief due to glide is
correctly reproduced and that the proper steady-states are achieved.
\paragraph*{}
At low values of $v_{ss}$ ($\lesssim0.06\gamma_{P}$), a more 
surprising type of motion
occurs in which the dislocation overcomes the Peierls barrier, glides a short
distance, and then comes to a stop. The cycle then repeats itself once enough
strain is re-accumulated to overcome the barrier again. This oscillatory motion 
will occur whenever the velocity assumed just above the Peierls barrier is 
greater than the theoretical $v_{ss}$ for the system. The rate at which
the dislocation glide relieves strain is temporarily greater than the applied
strain rate, so the energy falls below the Peierls barrier and glide is
no longer possible until the strain energy again increases sufficiently.

\subsubsection{Viscous Dynamics\label{section3b3}}
\begin{figure}
\includegraphics[width=88mm]{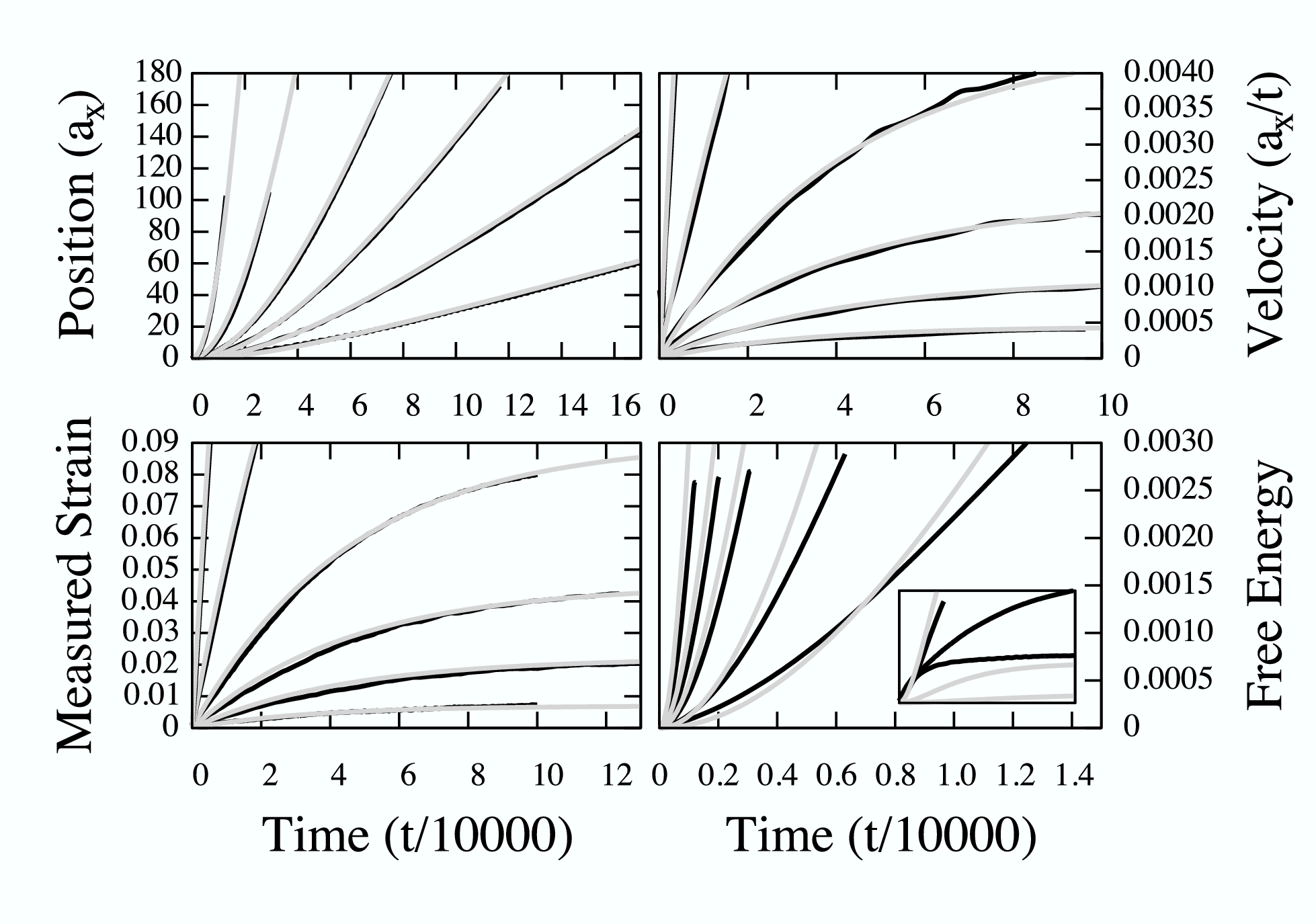}
\caption{
Additional comparisons between simulation data (black lines) and viscous motion 
equations (gray lines) for glide where 
where $M_{\gamma}$ is the only adjustable parameter.
From left to right in each plot curves are shown for $\dot{\gamma}=2\times 
10^{-5}$, $6\times 10^{-6}$, $2\times 10^{-6}$, $1\times 10^{-6}$, 
$5\times 10^{-7}$,
and $2\times 10^{-7}/t$ except for the lower right which shows data for
$\dot{\gamma}=2\times 10^{-4}$, $1.2\times 10^{-4}$, $8\times 10^{-5}$,
$4\times 10^{-5}$, and $2\times 10^{-5}/t$. The inset in the lower right corner
of the lower right plot shows data for lower shear rates, $\dot{\gamma}=
6\times 10^{-6}$, $2\times 10^{-6}$, and $5\times 10^{-7}/t$ where 
Eq. (\ref{energyt}) begins to fail. 
In all plots $r=-0.8$ and ($L_x,L_y$)=(56,46).
\label{gallvst2}
}
\end{figure}

\begin{figure}
\includegraphics[width=88mm]{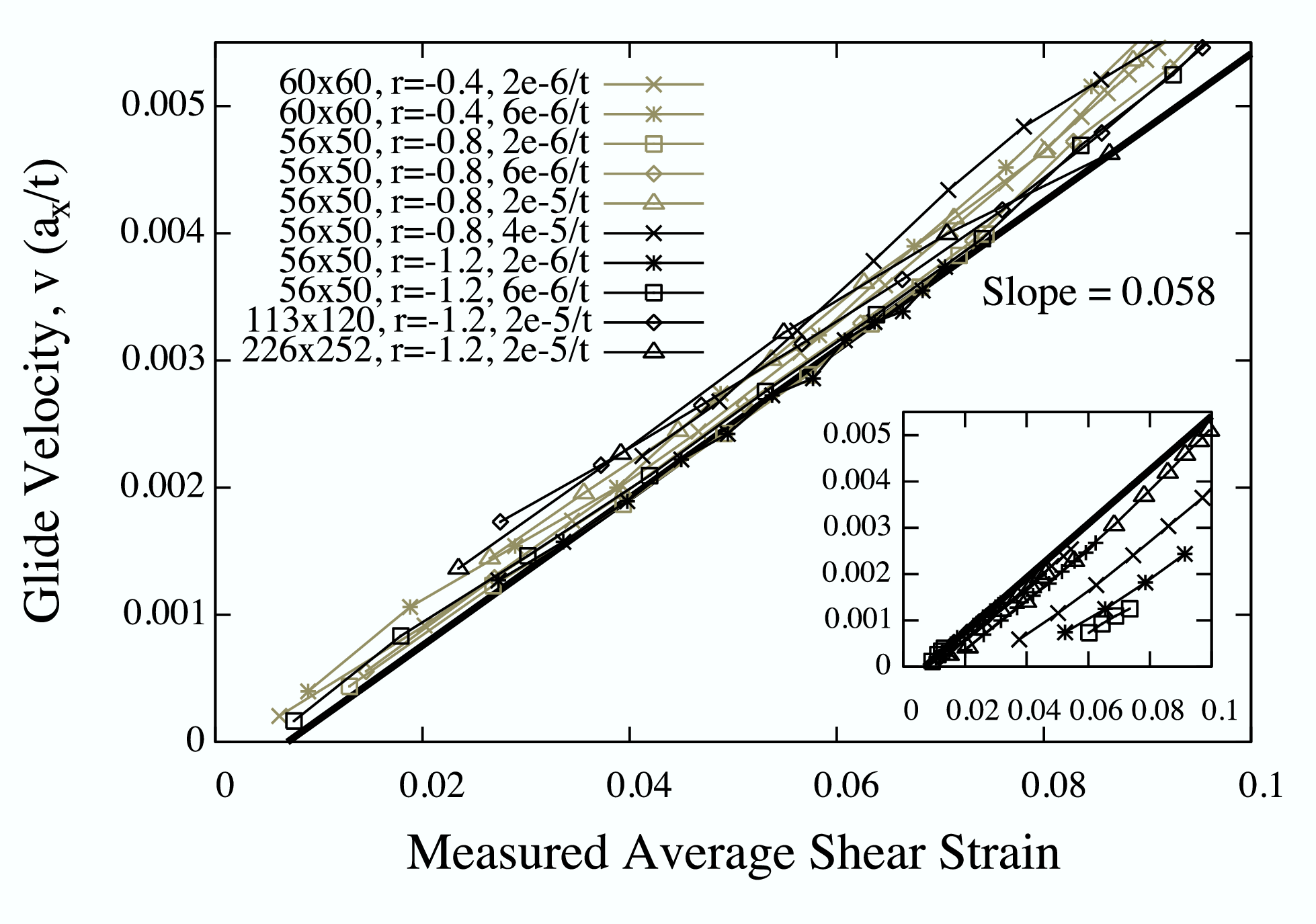}
\caption{
Dislocation glide velocity under rigid displacement as a function of the
measured average shear strain, $\bar{\gamma}$.
A number of system sizes, temperatures, and shear rates are shown to illustrate
the uniformity of the dynamics.
The heavy line is a representative linear fit. 
Inset: Dislocation glide velocity under relaxational displacement at
various shear rates and two values of $r$. The strain values are 
overestimated due to the nonlinear shear profile produced by this
type of shearing, but the slopes are relatively unchanged.
The heavy line is the same linear fit as in the larger graph.
\label{velvstr}
}
\end{figure}

\paragraph*{}
Empirically, dislocation glide velocity is described by the following equation:
\begin{equation}
v=v_s(\tau_{{\rm eff}}/\tau_s)^m
\label{gvel}
\end{equation}
where $v_s$ is the shear wave velocity, $\tau_{{\rm eff}}$ is the effective
shear stress on the dislocation, $\tau_s$ is the material stress constant,
and $m$ is the stress exponent \cite{dd90}. The stress exponent has been
found to range from less than 1 to over 100 in some cases. For typical pure
metals such as aluminum or copper, $m\simeq1$--5. These values may change
significantly depending on temperature, stress range, and local defect
densities. For example, in iron, $m$ falls into one of three regions 
({\it m}$<$1, $m$=1, {\it m}$>$1) depending on the conditions examined 
\cite{dd90}. 
As will be shown in this subsection, the dislocation velocity was found to be 
approximately linear in both stress and strain ($m\simeq 1$) for all
parameter ranges studied. This is not unexpected, as higher values of $m$ are
often attributed to effects such as jogs, impurities, and other defects which
modify the dynamics from those expected for pure, two-dimensional crystals. 
\paragraph*{}
The dynamics of a single gliding dislocation 
are well described by the 
equation of motion for a point mass in a damped medium;
\begin{equation}
m_{{\rm eff}}\dot{v}(t)=F_0-\beta v(t)
\label{newton}
\end{equation}
where $m_{{\rm eff}}$ is an effective dislocation mass, $F_0$ is a constant
proportional to $\dot{\gamma}$, and $\beta$ is a damping constant.
\paragraph*{}
Equations for $v(t)$, $x(t)$, and $\bar{\gamma}(t)$ 
can easily be derived from this
starting point, but first the Orowan equation will be used to write 
$m_{{\rm eff}}$, $F_0$, and $\beta$ in terms of more meaningful parameters.
It will be shown that the velocity is linear in $\bar{\gamma}$, 
but assuming this 
for now, one can write
\begin{equation}
v(\bar{\gamma})=M_{\gamma}(\bar{\gamma}(t)-\bar{\gamma}_0)
\label{linv}
\end{equation}
where $M_{\gamma}$ is the slope, which can be interpreted as an
effective mobility for glide.
Next note that $\bar{\gamma}(t)$ is a function of the applied strain 
and the strain relieved by the gliding dislocation;
\begin{equation}
\bar{\gamma}(t)=\dot{\gamma} t-\rho_d b x(t).
\label{straint}
\end{equation}
Substituting Eq.~(\ref{straint}) into Eq.~(\ref{linv}) and differentiating gives
\begin{equation}
\dot{v}(t)=M_{\gamma}\dot{\gamma}-M_{\gamma}\rho_d 
b v(t)
\label{vorowan}
\end{equation}
and equating terms in Eqs.~(\ref{vorowan}) and (\ref{newton}) shows that
\begin{equation}
\frac{F_0}{m_{{\rm eff}}}=M_{\gamma}\dot{\gamma}
\end{equation}
and
\begin{equation}
\frac{\beta}{m_{{\rm eff}}}=M_{\gamma}\rho_d b.
\label{beta}
\end{equation}
This analysis indicates that the damping experienced by the dislocation is
a result of the strain relief connected to the glide process and is not 
directly linked to the dynamics of Eq. (\ref{dyn}). That is, 
if the second term on 
the right hand side of Eq. (\ref{straint}) were removed then both the 
velocity and the strain would be linear functions of time, without any 
effective damping. Including this term means that the effective damping 
can be controlled by changing $\rho_d$, with larger values of $\rho_d$
corresponding to increased damping.
\paragraph*{}
Solving Eq. (\ref{newton}) in terms of these new parameters, and applying
the initial conditions $v(0)=0$ and $x(0)=0$ gives
\begin{equation}
v(t)=v_{ss}(1-e^{-M_{\gamma}\rho_d bt})
\label{velocity}
\end{equation}
and
\begin{equation}
x(t)=v_{ss}\bigg(t+\frac{1}{M_{\gamma}\rho_d b}
e^{-M_{\gamma}\rho_d bt}\bigg)-
\frac{\dot{\gamma}}{M_{\gamma}\rho_d^2 b^2}.
\label{position}
\end{equation}
Substituting Eq. (\ref{position}) into Eq. (\ref{straint}) then gives
\begin{equation}
\bar{\gamma}(t)=\frac{v_{ss}}{M_{\gamma}}
\big(1-e^{-M_{\gamma}\rho_d bt}\big)+\bar{\gamma}_0.
\label{straint2}
\end{equation}
Finally, comparing Eqs. (\ref{straint2}) and (\ref{velocity}) produces the
desired linear relation assumed in 
Eq. (\ref{linv}) and the similar relation 
$v_{ss}=M_{\gamma}\bar{\gamma}_{ss}$, where
$\bar{\gamma}_{ss}=\dot{\gamma}/M_{\gamma} \rho_d b$. 
The data shown in Figs. 
\ref{velvstr} and \ref{velss} verify that these linear relations 
are observed.
\paragraph*{}
In all of these equations, the only adjustable parameter is 
$M_{\gamma}$, 
the effective
mobility of the dislocation. Using values of $M_{\gamma}$ 
measured from simulations,
Fig.~\ref{gallvst} shows excellent agreement between these 
analytic results and the simulation data for one parameter set, and Fig.
\ref{gallvst2} shows similar agreement for various other parameter sets. 
If it is assumed that the free energy obeys the relation to $\bar{\gamma}$ 
given in Eq. (\ref{dFrigid}), then 
Eq. (\ref{straint2}) can be substituted into Eq. (\ref{dFrigid}) to give
\begin{equation}
\Delta F_{Shear}=\frac{1}{6}\bigg[\frac{q^2_{eq}A \dot{\gamma}}
{M_{\gamma}\rho_d b}\big(1-e^{-M_{\gamma}\rho_d bt}\big)\bigg]^2
\label{energyt}
\end{equation}
which agrees relatively well with the high shear rate data, as shown in Fig. 
\ref{gallvst2}. The inset in the lower right of Fig. 
\ref{gallvst2} shows how the agreement begins to fail at lower shear rates.
This anomaly in the low $\dot{\gamma}$ glide data is not fully 
understood.
\paragraph*{}
It is worth examining the strain dependence of the velocity further.
In gradient systems, the velocity of finite structures is expected to 
be proportional to the driving force applied $F_D$, which in this case
can be interpreted as the derivative of the change in free energy due 
to the application of shear; 
\begin{equation}
v\sim F_D=
\frac{d\Delta F_{Shear}}{d\bar{\gamma}}\simeq
\frac{q_{eq}^4 A^2}{3}\bar{\gamma}. 
\label{velforce}
\end{equation}
Additionally, Eqs. (\ref{newton})--(\ref{straint2}) indicate that velocity 
is in general linear in $\bar{\gamma}$ for this type of overdamped system.
All simulations resulted in approximately 
linear velocity ($v$) vs. $\bar{\gamma}$ behavior for dislocation glide, as 
shown in Fig.~\ref{velvstr}.
It is important to correct the 
overall strain shown in Fig.~\ref{velvstr} for that relieved by the glide of 
the dislocation (Eq. (\ref{orowan})), especially when using small system sizes.
\paragraph*{}
Both methods of displacement produce nearly the same value of 
$M_{\gamma}$ under all conditions, though it is more difficult
to determine the local strain around the dislocation for the case of
relaxational displacement, due to the nonlinear shear profile.
For rigid displacement, the free energy follows the expected form 
($\Delta F_{Shear}\sim\bar{\gamma}^2$) and the velocity
appears to be linear for $\bar{\gamma}$ less than $\sim10$\%.
For relaxational displacement, the anomaly in the free energy behavior
noted above complicates the results,
but the velocity remains linear in $\bar{\gamma}$ with 
values of $M_{\gamma}$ similar to those found for rigid
displacement.
An analytic calculation of $M_{\gamma}$ would complete this 
analysis of the dynamics, but since the simulation results indicate no
strong dependencies on any variables, $M_{\gamma}=0.06a_x/t$ is
believed to be a reasonable estimate for most cases of interest.
\paragraph*{}
Shear was also applied along directions not lying on one of the axes of symmetry
with predictable results. As the angle $\theta_R$ is increased 
(with $0^{\circ}$ denoting 
alignment with a symmetry axis), the Peierls barrier grows but the slip 
direction
remains along the nearest symmetry axis. Once $\theta_R$ becomes large enough, 
approximately $10-30^{\circ}$ depending on the value of $r$, the 
dislocation prefers
to climb rather than glide, with motion in the general direction of the 
applied shear.
\paragraph*{}
A similar analysis to that presented in Eqs. (\ref{newton})--(\ref{energyt}) 
can be applied to the case of constant strain by 
removing the external force from Eq.~(\ref{newton}). The resulting equations
were also found to agree well with simulation data.
It is also worth noting that the velocity vs. $\bar{\gamma}$ behavior is
essentially the
same as that shown in Fig.~\ref{velvstr} when the shear condition is one of 
constant strain.

\subsection{Climb: Constant Applied Strain Rate Dynamics\label{section3c}}
\begin{figure}
\includegraphics[width=80mm]{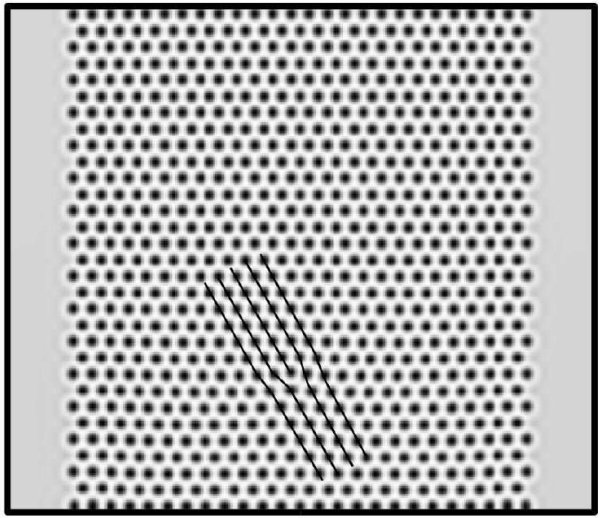}
\caption{
A small sample climb simulation setup where $\rho(x,y)$ has been plotted.
The extra row of particles terminating at the core of the dislocation has been
highlighted.
\label{climbgif}
}
\end{figure}

\paragraph*{}
Climb simulations were conducted using steady compression over a range of
parameter values similar to those used for glide simulations. 
Before presenting the results, a caveat on this portion of the study is in 
order. It was found that the results varied systematically (i.e. the 
Peierls barrier decreased) with the grid spacing $\Delta x$, apparently 
due to the
decrease in relevant dimensions with compression. A grid spacing small enough 
to overcome this effect could not be reached since the time step must be 
dramatically decreased with $\Delta x$. But the nature of the results and the 
essential physics remain the same; the data is only shifted by this effect.
An example of the climb simulation geometry is shown in Fig. \ref{climbgif}.

\subsubsection{Peierls Barrier for Climb\label{section3c1}}
\begin{figure}
\includegraphics[width=88mm]{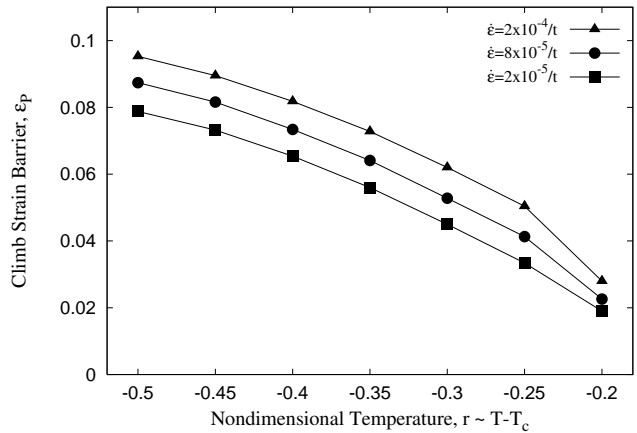}
\caption{
Temperature dependence of the Peierls strain barrier for climb 
without thermal fluctuations.
Data shown is at $\rho_0=0.25$ and ($L_x,L_y$)=(52,103) 
under rigid displacement.
\label{peivtempc}
}
\end{figure}

\begin{figure}
\includegraphics[width=88mm]{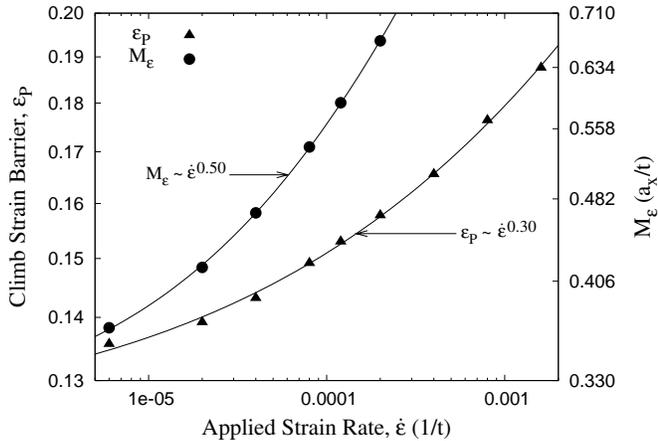}
\caption{
Measured Peierls strain barrier for climb 
and the subsequent mobilities
under rigid displacement at $r=-1.2$ and ($L_x,L_y$)=($52,166$).
\label{peiBvratec}
}
\end{figure}

\paragraph*{}
The dependence of $\epsilon_{P}$ on $\rho_d$ is of the same nature as 
that found for $\gamma_{P}$. No change was found under rigid 
displacement as $L_x$ and $L_y$ were increased, but an increase with $L_x$ was
observed under relaxational displacement, again in proportion to the 
diffusion time from the edge of the sample to the dislocation core.
\paragraph*{}
The $r$ dependence of $\epsilon_{P}$ is shown in Fig. \ref{peivtempc}
for various strain rates. Comparison with the glide Peierls
barrier data in Fig. \ref{peivrate} confirms the same 
general linear behavior. 
$\epsilon_{P}$ is quite large at low $r$ but decreases 
toward $T_c$ such that there is a crossover close to $T_c$ where 
$\epsilon_{P}$
becomes less than $\gamma_{P}$.
Thus climb is predominant at high temperatures, in agreement with the 
accepted phenomenology \cite{dd90}. This was also confirmed in the
glide simulations where climb was found be preferred near $T_c$, even at
very low values of applied shear.
Note that the data shown in Fig. \ref{peivtempc} was obtained using modified
boundary conditions of mirror on all sides with no liquid phase.
\paragraph*{}
Following \cite{ken04}, the change in $F$ under compression can be calculated
by substituting $\rho(x/(1+\bar{\epsilon}),y)$ into 
Eq. (\ref{free}) and minimizing
with respect to $A$. The result is similar to that for shear;
\begin{equation}
\Delta F_{Comp.}=\frac{q^4_{eq}A^2}{2}\bar{\epsilon}^2.
\label{dFcomp}
\end{equation}
In this limit, as was also the case for glide, $\epsilon_{P}$ 
can be written in the form
\begin{equation}
\epsilon_{P}=\sqrt{\frac{2\Delta F_{P}^{Climb}(\rho_0,r)}{q^4_{eq}A^2}}.
\label{peirigidc}
\end{equation}
\paragraph*{}
The strain rate dependence is also similar to that for glide, as shown more
clearly in Fig. \ref{peiBvratec}. 
The results show that $\gamma_{P}\sim \dot{\gamma}^{0.30}$, which is 
similar to the dependence $\gamma_{P}\sim \dot{\gamma}^{0.37}$ 
measured for glide at the same $r$.
The absolute values of $\epsilon_{P}$ are significantly higher than those
for glide in this case because of the low value of $r$ that was used.
The same arguments leading to Eq. (\ref{hoptime3}) should apply to 
climb as well, since the behavior seems to be essentially the same as
that found for glide.

\subsubsection{Atomistic Climb Mechanism\label{section3c2}}
\begin{figure}
\includegraphics[width=84mm]{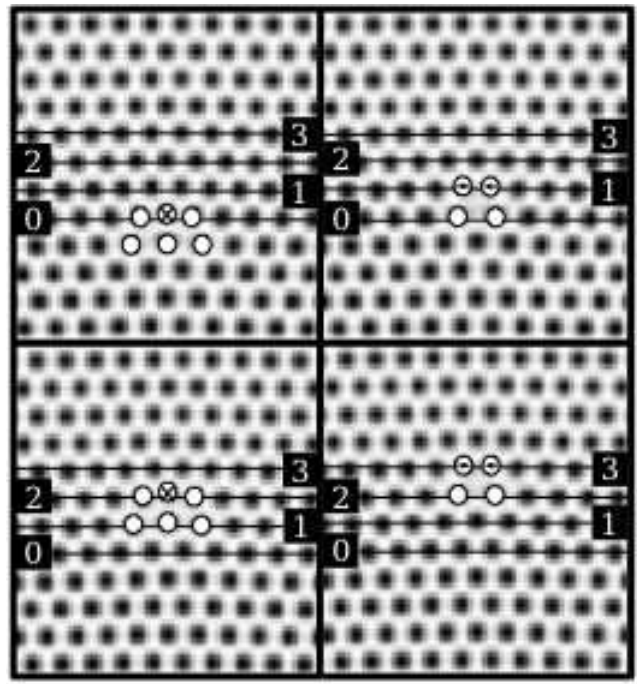}
\caption{
Atomistic climb mechanism under constant applied strain rate. 
From top left to bottom right, $\rho(x,y)$ is shown at $t$=300, 600, 800, and
900, corresponding to $\bar{\epsilon}$=2.4$\%$, 4.8$\%$, 6.4$\%$, and 7.2$\%$.
The particles
around the dislocation core have been highlighted and the rows near the 
core have been labeled for clarity.
The particles marked with an 'X' are those which diffuse away between
subsequent images, and those marked with arrows merge together.
\label{climbmech}
}
\end{figure}

\paragraph*{}
Dislocation climb is a nonconservative process. It requires either the diffusion
of particles away from the dislocation core or toward it, unlike glide which 
involves only rearrangements of particles around the core. The mechanism of
climb is shown in Fig. \ref{climbmech}, where in these simulations mass
diffuses away from the core since the strain is applied through compression. 
\paragraph*{}
Again, similar to what was found for glide, the motion has a stick-slip 
character at low velocities and becomes more continuous at higher velocities.
The motion proceeds by alternating between configurations 1 and 2 (Fig. 
\ref{configs}). Starting from Config. 2, as shown in the upper left image of
Fig. \ref{climbmech}, the particle marked with an 'X' diffuses away, leaving
the core in Config. 1 as shown in the next image. The two particles marked with
arrows then merge together, returning the dislocation to Config. 2 as shown in
the subsequent image. 
The process repeats as long as there is sufficient strain
energy to maintain motion. 
For climb in the opposite direction, particles 
diffuse in and split rather then diffuse away and merge, respectively. 
This merging and splitting of particles may seem unphysical, 
but in a time-averaged sense these motions simply represent 
diffusion of mass away from or toward the dislocation core, 
which is the fundamental limiting process in dislocation climb.

\subsubsection{Viscous Dynamics\label{section3c3}}
\begin{figure}
\includegraphics[width=84mm]{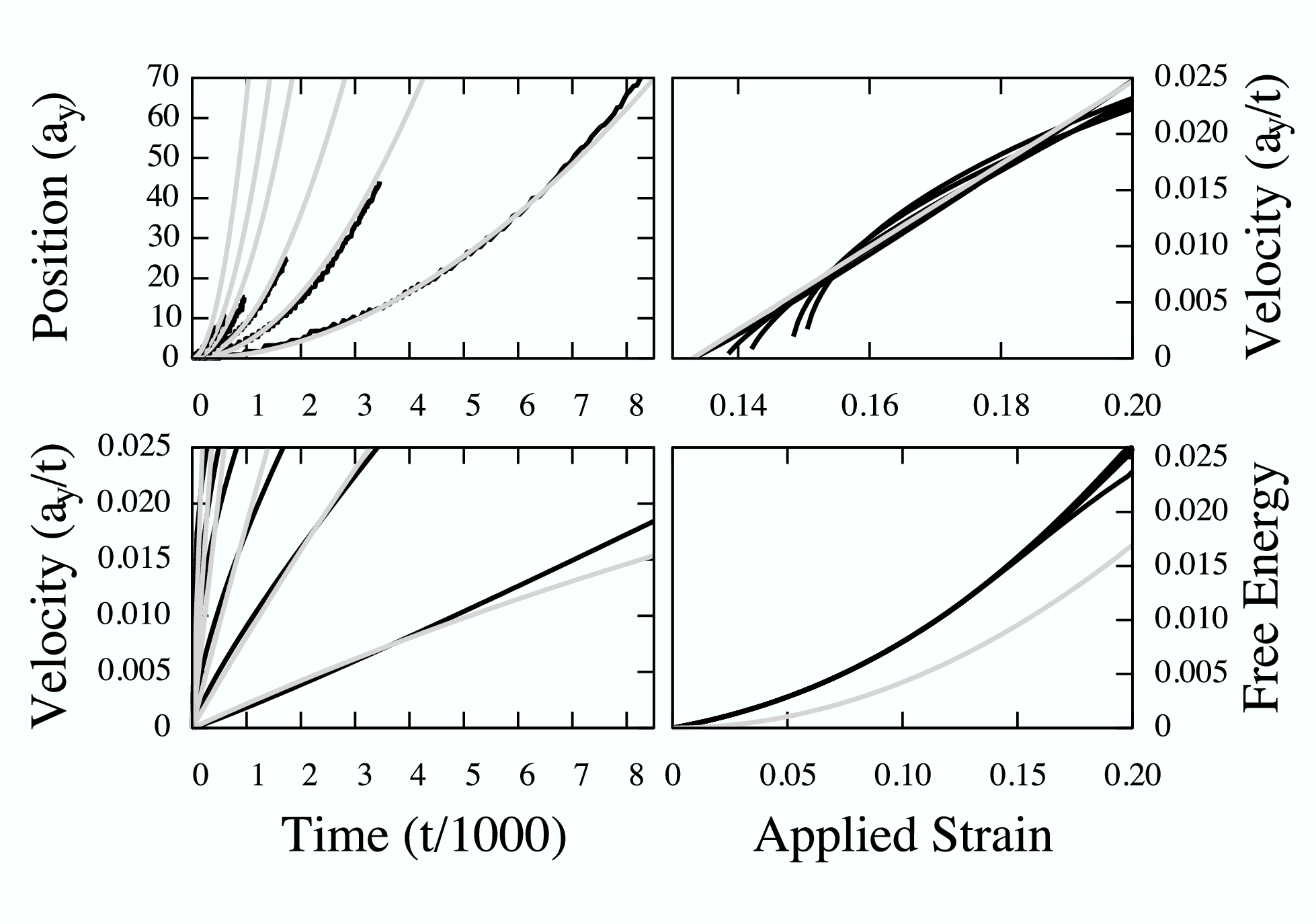}
\caption{
Comparisons between simulation data (black lines) and viscous motion 
equations (gray lines) for climb.
From left to right in each plot curves are shown for $\dot{\epsilon}=2\times 
10^{-4}$, $1.2\times 10^{-4}$, $8\times 10^{-5}$, $4\times 10^{-5}$, 
$2\times 10^{-5}$,
and $6\times 10^{-6}/t$.
In all plots $r=-1.2$ and ($L_x,L_y$)=(52,166).
\label{callvst2}
}
\end{figure}

\paragraph*{}
The dynamics of a single climbing dislocation are well described by the same 
damped equation of motion used to describe glide (Eq. (\ref{newton})).
Again, the only adjustable parameter is $M_{\epsilon}$, 
the effective mobility for dislocation climb.
Fig.~\ref{callvst2} shows the agreement between these 
analytic results and typical sets of simulation data.
\paragraph*{}
The velocity versus $\bar{\epsilon}$ behavior shown in Fig. \ref{callvst2} 
appears to be slightly 
nonlinear, but this is due to the relatively short range of motion that
could be captured with computationally tractable system sizes.  
An approximate $M_{\epsilon}$ can nonetheless be extracted, and the results
indicate first of all that
the values of $M_{\epsilon}$ are an order of magnitude 
higher than those measured for $M_{\gamma}$ ($M_{\epsilon}\simeq0.5$). 
The slopes of the
$v$ versus $\bar{\epsilon}$ curves are much steeper for climb than for glide, 
but at the same time the velocities remain zero to much higher strains due
to the larger values of $\epsilon_{P}$ (except near $T_c$).
Also, $M_{\epsilon}$ is not quite as unchanging as $M_{\gamma}$, in that
relatively weak, though measurable dependencies on $r$ and 
$\dot{\epsilon}$ were found. 
The data indicates a slight decrease in $M_{\epsilon}$ with increasing $r$
and an increase with $\dot{\epsilon}$ that goes like $\sqrt{\dot{\epsilon}}$
(Fig. \ref{peiBvratec}).
\paragraph*{}
To calculate the dynamics of $F$, Eq. (\ref{straint2}) can be substituted 
into Eq. (\ref{dFcomp}) to give
\begin{equation}
\Delta F_{Comp.}=\frac{1}{2}\bigg[\frac{q^2_{eq}A \dot{\epsilon}}
{M_{\epsilon}\rho_d b}\big(1-e^{-M_{\epsilon}\rho_d bt}\big)\bigg]^2
\label{energytcomp}
\end{equation}
which agrees reasonably well with the data shown in Fig. \ref{callvst2}. 
The difference is mostly due to the low value of $r$ used, since the 
one mode approximation loses accuracy away from $T_c$.
No anomaly in $F$ like that found in the glide data was observed in the
climb simulations. All curves of the change in $F$ under compression
fall onto approximately the same curve when plotted versus $\bar{\epsilon}$. 
\paragraph*{}
Compression was also applied along directions not lying on one of the axes of 
symmetry at $r=-0.8$. As the angle $\theta_R$ is increased, the dislocation 
first glides
some distance proportional to $\theta_R$ in a direction along the nearest 
symmetry axis. Then climb begins along the same lattice direction as in the 
unrotated case, with the value of $\epsilon_{P}$ increasing only slightly
with $\theta_R$. Nearer $T_c$ it would be reasonable to expect less tendency
toward the initial gliding, as climb becomes the preferred type of
motion. Generally speaking, the application of strain along irregular directions
relative to the lattice symmetry results in a mixed motion of glide and climb.

\subsection{Annihilation\label{section3d}}
\begin{figure}
\centering
\includegraphics[width=84mm]{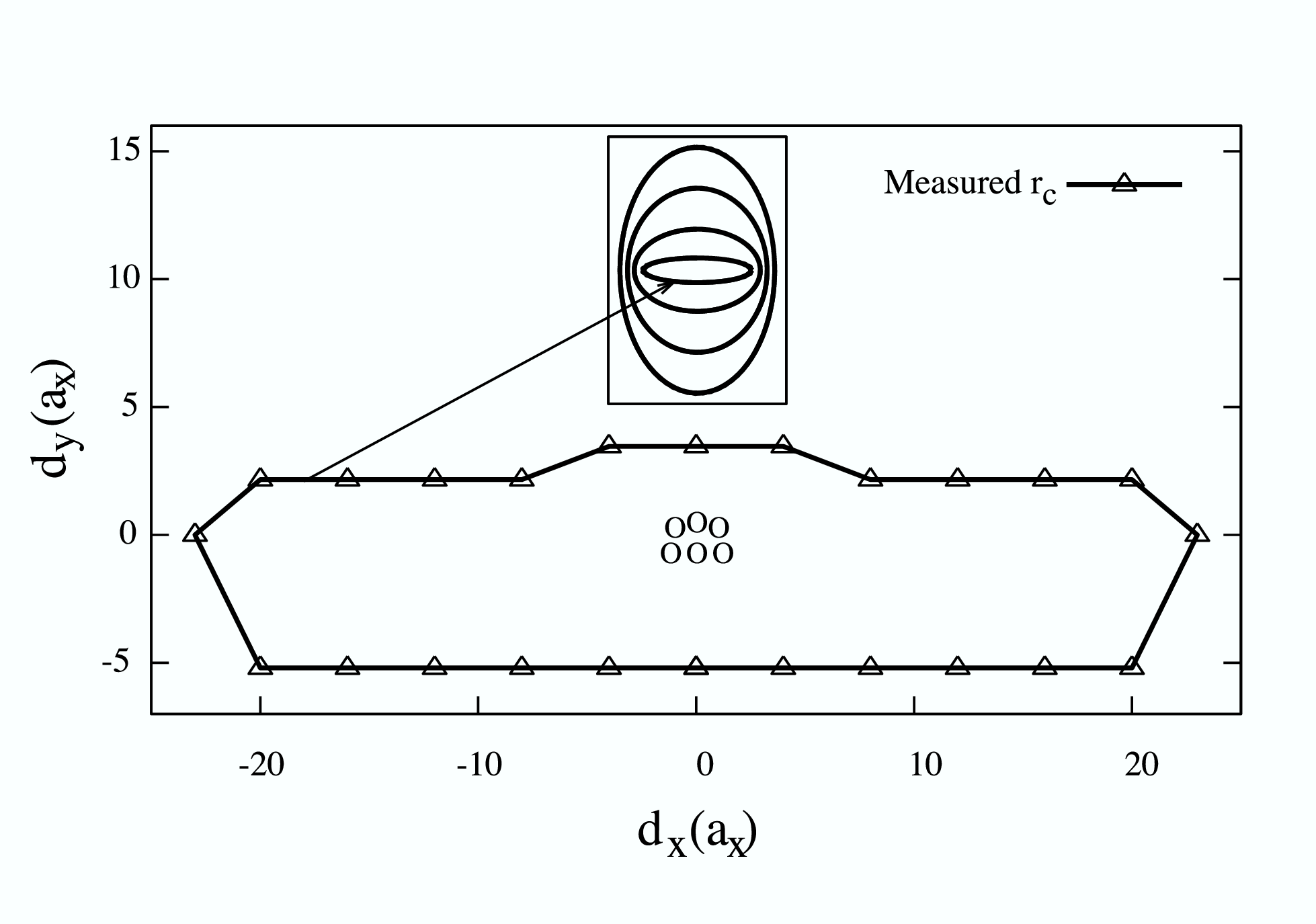}
\caption{
Measured critical radii for annihilation at $r=-0.25$ and $\rho_0=0.25$.
The configuration shown in the center is the reference dislocation at
$(0,0)$ from which $\theta_0$ was measured.
The inset shows a schematic of the expected behavior as the temperature
is increased above the crossover $r$ at which climb becomes dominant.
\label{annihrc}
}
\end{figure}

\paragraph*{}
Annihilation occurs when two dislocations having opposite burger's
vectors merge and eliminate each other. There exists a critical separation,
$r_c$, at a given angle, $\theta_0$, below which annihilation will occur, 
and this 
separation is in principle a function of the crystal symmetry, type of 
dislocation, temperature, relative velocity, and
the local strain field. Results were obtained here for the static case ($v=0$)
at a single temperature and under no applied strain, for two perfect 
edge dislocations. 
\paragraph*{}
Consider one dislocation at some location $(0,0)$ and another at 
$(d_x,d_y)$ with opposite burger's vector. In radial coordinates the
separation can be expressed in terms of a distance $r_0$ and an angle 
$\theta_0$.
At $\theta_0=0^{\circ}$, annihilation occurs by pure glide, and as $\theta_0$ 
is increased a mixed motion of glide and climb is required, until 
$\theta_0=90^{\circ}$ where annihilation occurs by pure climb.
$r_c$ was determined as a function of $\theta_0$
by increasing the initial separation until annihilation 
no longer occurred. Periodic boundary conditions were used in all directions and
the parameters chosen were $\rho_0=0.25$, $r=-0.25$, and $(L_x,L_y)=(56,43)$.
The equilibrium wavenumber at this $\rho_0$ and $r$ would require 56.5 
particles in the $x-$direction, so placing 56/row in the bottom half and
57/row in the top half produces a dislocation with no preset bias toward 
climb in either direction.
The results are shown in Fig. \ref{annihrc}.
\paragraph*{}
Despite the unbiasing, $r_c$ is asymmetric with a preference toward climb 
in the $-y$ direction. This is apparently a consequence of the asymmetry
of the strain field across the $x-$axis of the dislocation core,
where there is an enhancement of strain in the lower half-plane. 
The particle positions around the core clearly reflect this asymmetry.
Note that the details of the strain field will be slightly different for a 
dislocation in Config. 1, but the same argument should nonetheless hold.
\paragraph*{}
The elliptical shape of $r_c(\theta_0)$ is expected since 
$\gamma_{P}^0<\epsilon_{P}^0$ for this parameter set. As $r$ is increased,
eventually $\epsilon_{P}^0<\gamma_{P}^0$, and the primary axis of the ellipse
should coincide with the $y-$axis (climb axis), becoming more elliptical as
$T_c$ is approached. This expected behavior is shown schematically in the 
inset of Fig. \ref{annihrc}. Moving from the inner to the outer ellipse 
corresponds to increasing $r$.
\paragraph*{}
Extending the elliptical approximation and assuming that $r_c$ is directly
proportional to the Peierls strain, one can write a temperature dependent
equation for $r_c$;
\begin{equation}
r_c(\theta_0,r)\simeq\frac{|A_{\gamma}A_{\epsilon}r|}
{\sqrt{A_{\gamma}^2 \sin^2{\theta_0}+A_{\epsilon}^2 \cos^2{\theta_0}}}
\label{anniheq}
\end{equation}
where $A_{\gamma}$ and $A_{\epsilon}$ are the slopes of the Peierls strain 
versus $r$ curves for glide and climb respectively.

\section{Conclusion\label{conclusion}}
\label{sec:level4}
\paragraph*{}
Three fundamental dislocation processes have been numerically examined in 
idealized two dimensional settings using a phenomenological PFC model.
The diffusive dynamics were measured over a range of temperatures, dislocation
densities, and experimentally accessible strain rates.
In equilibrium, two stable edge
dislocation configurations were found to exist, with one resulting in a 
slightly lower Peierls barrier for glide than the other. 
The Peierls barriers for glide and climb, $\gamma_{P}$ and $\epsilon_{P}$
respectively, were found
to have little or no dependence on dislocation density, and both showed
approximately linear decreases with increasing temperature (in the
absence of thermal fluctuations). Near $T_c$, $\epsilon_{P}<\gamma_{P}$
verifying the expectation that climb is dominant at high temperatures.
A crossover temperature was identified below which $\gamma_{P}<\epsilon_{P}$
and glide becomes the preferred type of motion. 
Both strain barriers also showed essentially power law increases with the
applied strain rate, where the exponents are similar for glide and 
climb at equal $r$'s. Under relaxational displacement (no phonons), 
$\gamma_{P}$ is nearly linear in $\dot{\gamma}$ and goes as 
Eq. (\ref{hoptime3}), while under rigid 
displacement at high strain rates (strong phonons) the deviation from linear
is much greater ($\gamma_{P}\simeq \dot{\gamma}^{0.38}$) with relatively 
little change in the barrier strain at high strain rates.
Physical arguments and some mathematical arguments were given for all of 
these behaviors.
\paragraph*{}
Rigid displacement with $\dot{\gamma},\dot{\epsilon}>5\times10^{-6}/t$
most accurately reproduces the rigid behavior of a real crystal.
A more rigorous PFC model, derived from microscopics, that will include a wave 
term to simulate phonon dynamics is currently being developed \cite{mpfc}. 
Based on the results presented here, it is expected that
this model will produce dynamics that fall between the limits of 
relaxational and rigid response.
At strain rates below approximately $2\times10^{-7}/t$,
the two methods of displacement are essentially equivalent.
\paragraph*{}
The motion of a gliding or climbing edge dislocation was found to be 
stick-slip in character at low velocities
and nearly continuous at high velocities. Three possible regimes of motion were
observed for glide, 
depending on the expected steady-state velocity of the dislocation
defined in Eq.~(\ref{vss}).
These involve an oscillatory glide, a steady-state glide, and 
slipping rows of particles, in order of increasing $v_{ss}$.
\paragraph*{}
A simple viscous dynamic model has been formulated to describe the results 
obtained for gliding and climbing dislocations, where the only adjustable
parameter is $M_{\gamma}$ or $M_{\epsilon}$. Excellent agreement is obtained
between these equations and the simulation results, both of which 
indicate that velocity is linear in strain for both glide and climb. 
The slope of the $v$ versus $\bar{\gamma}$ curve for glide, $M_{\gamma}$
was found to be nearly unchanging across all parameter ranges.
The slope for climb, $M_{\epsilon}$, 
which is an order of magnitude greater than $M_{\gamma}$, 
was found to increase approximately as $\sqrt{\dot{\epsilon}}$.
\paragraph*{}
A critical distance for the annihilation of two edge dislocations was also
measured, and an asymmetry with preference toward annihilation in the
$-y$ direction was found. $r_c(\theta_0)$ approximately takes the form of an
ellipse whose major axis is predicted to be 
along the glide direction at low temperatures
and along the climb direction at high temperatures.
\paragraph*{}
Other phase-field models have recently been used to study dislocation dynamics
\cite{cuitino01,koslowski02,wang01b,wang05}. 
These approaches differ from the PFC method in that they do not naturally
contain atomistic detail.
The domains in these models typically differentiate dislocation loops 
and the interfaces represent dislocation lines.
Coarsening of large arrays of lines etc. can be efficiently
studied, but atomistic detail is either lost or must be explicitly
added through postulated Peierls potentials. The relevant equations of 
elasticity must also be rigorously applied, unlike in the PFC model which
naturally exhibits elastic behavior as well as Peierls potentials.
\paragraph*{}
Other phenomena relevant to 
dislocation dynamics, such as obstacle and impurity effects, 
could be studied with a similar approach,
and more complicated dynamics involving screw dislocations, dislocation
loops, multiplication processes, etc. could be examined in three dimensional 
simulations. Alternatively, the two dimensional model could provide interesting
insights into the problem of dislocation-mediated melting in two dimensions.

\begin{acknowledgments}
JB would like to acknowledge support from a Richard H. Tomlinson Fellowship, 
KRE acknowledges support from the NSF under Grant No. DMR-0413062,
and MG was supported by the Natural Sciences and Engineering Research 
Council of Canada and by {\it le Fonds Qu\'eb\'ecois de la recherche sur
la nature et les technologies}.
\end{acknowledgments}


\end{document}